\documentclass[showpacs,preprintnumbers,amsmath,amssymb,prb]{revtex4}

\def\XXint#1#2#3{{\setbox0=\hbox{$#1{#2#3}{\int}$}
     \vcenter{\hbox{$#2#3$}}\kern-.5\wd0}}

\usepackage{graphicx}
\usepackage{subfigure}
\usepackage{bm}

\begin{document}

\title{Low-energy singlet sector in spin-$\frac12$ $J_1$--$J_2$ Heisenberg model on square lattice}

\author{A.\ Yu.\ Aktersky$^{1}$}
\email{aktersky@gmail.com}
\author{A.\ V.\ Syromyatnikov$^{1,2}$}
\email{asyromyatnikov@yandex.ru}
\affiliation{$^1$National Research Center "Kurchatov Institute" B.P.\ Konstantinov Petersburg Nuclear Physics Institute, Gatchina 188300, Russia}
\affiliation{$^2$St.\ Petersburg State University, 7/9 Universitetskaya nab., St.\ Petersburg, 199034
Russia}

\date{\today}

\begin{abstract}

Based on a special variant of plaquette expansion, an operator is constructed whose eigenvalues give the low-energy singlet spectrum of spin-$\frac12$ Heisenberg antiferromagnet on square lattice with nearest- and frustrating next-nearest-neighbor exchange couplings $J_1$ and $J_2$. It is well known that a non-magnetic phase arises in this model at $0.4\alt J_2/J_1\alt 0.6$ sandwiched by two N\'eel ordered phases. In agreement with previous results, we observe a first-order quantum phase transition (QPT) at $J_2\approx 0.64J_1$ from the non-magnetic phase to the N\'eel one. Large gap ($\agt0.4J_1$) is found in the singlet spectrum at $J_2<0.64J_1$ that excludes a gapless spin-liquid state at $0.4\alt J_2/J_1\alt 0.6$ and the deconfined quantum criticality scenario for the QPT to another N\'eel phase. We observe a first-order QPT at $J_2\approx 0.55J_1$ presumably between two non-magnetic phases.

\end{abstract}


\maketitle

\section{Introduction}

Frustrated quantum antiferromagnets have provided a convenient playground for the investigation of novel types of many-body phenomena including quantum spin-liquid and nematic phases, novel universality classes of phase transitions, order-by-disorder phenomena, to mention but a few. One of the canonical models in this field is the spin-$\frac12$ Heisenberg antiferromagnet (AF) on square lattice with nearest- and frustrating next-nearest-neighbor AF exchange couplings $J_1$ and $J_2$ ($J_1$--$J_2$ model) whose Hamiltonian has the form
\begin{equation}
\label{ham}
{\cal H} = \sum_{\langle i,j \rangle}	{\bf S}_i{\bf S}_j 
+ 
J_2\sum_{\langle \langle i,j \rangle \rangle}	{\bf S}_i{\bf S}_j,
\end{equation}
where we put $J_1=1$. Despite a great deal of interest which this model has attracted during the last quarter of a century, its properties remain a puzzle in the most frustrated regime of $0.4\alt J_2\alt 0.6$.

This problem has been attacked using many powerful numerical and analytical methods including variational Monte-Carlo (VMC) calculations, \cite{vmc,vmcjap} density matrix renormalization group (DMRG) method, \cite{balents,fisher} coupled cluster method (CCM), \cite{darradi,richter} plaquette \cite{serplaq,serplaq2} and dimer \cite{serdim1,serdim2,serdim3} series expansions, tensor network state approach, \cite{tensnet,tensnet2} exact diagonalization, \cite{ed,ed2,ed3} spin-wave analysis, \cite{sw} large-$N$ expansion, \cite{sach1,sach2} hierarchical mean-field approach, \cite{mf} bond operator approach, \cite{kotov,bond,bond2} functional renormalization group, \cite{wolfle} and some others. It is generally believed that N\'eel ordered phases with AF vectors $(\pi,\pi)$ and $(0,\pi)$ (or $(\pi,0)$) arise at $J_2\alt0.4$ and $J_2\agt0.6$, respectively. Properties are still actively debated now of the magnetically disordered phase in the intermediate region of $0.4\alt J_2\alt 0.6$. Among definite conclusions about the nature of this phase are columnar dimer state, \cite{serplaq,sach2} gapped \cite{balents} and gapless \cite{vmc} spin liquids. However considerable amount of works reports plaquette valence bond solid (VBS) states \cite{ed3,fisher,tensnet2,mf,bond2} or a sufficiently close proximity of the system to such states \cite{serplaq2,ed2}.

Quantum phase transitions (QPTs) from the paramagnetic phase to N\'eel ones are of particular interest. A first-order QPT is what one can expect at $J_2\approx0.6$ in the Landau-Wilson paradigm of phase transitions due to different broken symmetries in two phases which cannot be connected by a group-subgroup relation. In agreement with this expectation, the majority of works report the first-order QPT at $J_2\approx0.6$ (see, e.g., Refs.~\cite{serplaq,tensnet,serdim2,tensnet2,balents,fisher,mf}). 

In contrast, QPT at $J_2\sim0.4$ remains a subject of controversy. Within the Landau-Wilson paradigm, one expects a first-order QPT if $C_4$ lattice rotational symmetry is broken in the paramagnetic phase \cite{mf} (e.g., in the case of columnar VBS). 
\footnote{
If $C_4$ lattice rotational symmetry is unbroken in the paramagnetic phase, one expects a second-order QPT within the Landau-Wilson paradigm. \cite{mf}
}
It is suggested, however, in Refs.~\cite{deconf,deconf2} that a second-order QPT can happen in this case via a mechanism of deconfined quantum criticality. Series expansion calculations of magnetic susceptibilities over different perturbation fields suggest the first-order QPT, \cite{serdim3} while calculation of the same susceptibilities by CCM shows the second-order QPT. \cite{darradi} Recent numerical works obtain an intrinsic structure of the nonmagnetic phase: a disordered phase with gapless triplet excitations at $0.4\alt J_2\alt 0.5$ (Refs.~\cite{vmcjap,richter,fisher}) and (plaquette \cite{fisher}) VBS phase with gapped triplet excitations at $0.5\alt J_2\alt 0.6$ (Refs.~\cite{vmcjap,fisher,tensnet}). Although authors do not exclude the deconfined criticality scenario at a single point or at a wide critical region around $J_2\sim0.5$ (quite definite conclusion about its existence can be found in Ref.~\cite{tensnet}), some of them point out that the system size available now for numerical computation is insufficient to make up a definite conclusion about the nature of the QPT (see, e.g., discussion in Refs.~\cite{vmcjap,fisher}).

\begin{figure}
\includegraphics[scale=0.5]{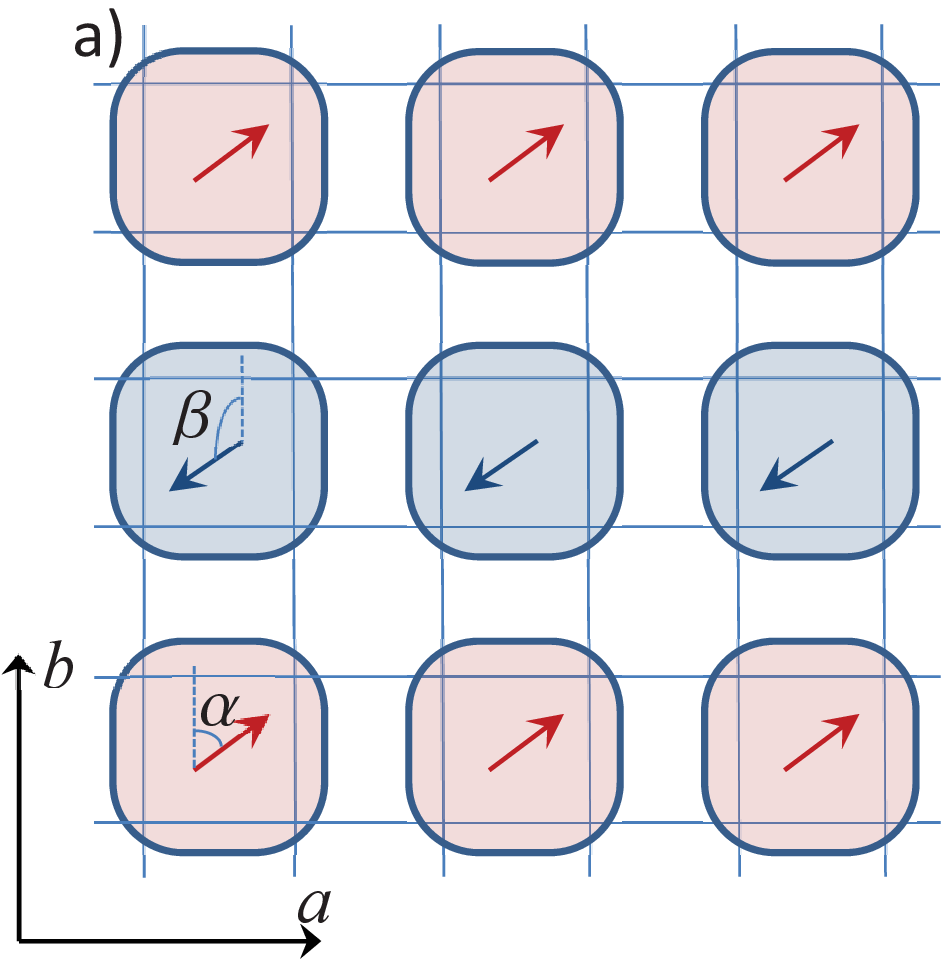}
\includegraphics[scale=0.5]{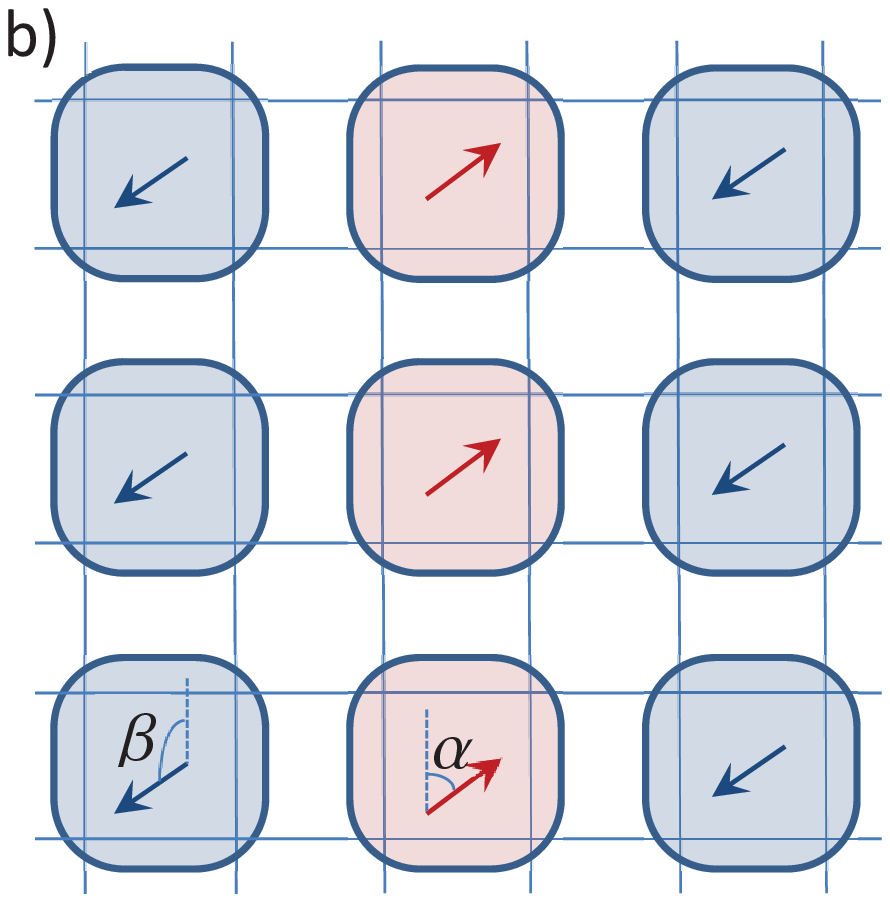}
\caption{(Color online.) Sketch of wave functions of the degenerate ground state with columnar plaquette structure proposed in the present paper at $0.55<J_2<0.64$. Singlet states of each plaquette are described in terms of states of pseudospin-$\frac12$. Orientation of pseudospins in $xz$ plane of the pseudospin space is depicted by arrows (dashed lines denote orientation of $z$ axis in the pseudospin space).
\label{gs}}
\end{figure}

Using an approach suggested in our previous paper \cite{singlon} which describes the low-energy singlet sector of model \eqref{ham}, 
\footnote{
Notice that the ground state of model \eqref{ham} was found to be singlet in previous studies while a rigorous proof (Marshall's theorem \cite{marsh,marsh2,auer}) of its singlet nature exists only for $J_2=0$ and $J_2\to\infty$. 
}
we demonstrate below that a first-order QPT arises at $J_2\approx 0.55$ presumably from the plaquette VBS phase to a plaquette phase having columnar structure (see Fig.~\ref{gs}). We find that ground state is separated by a large gap $\agt0.4$ from the first excited singlet level at $J_2<0.64$. Then, critical fluctuations cannot arise in the singlet sector that excludes the deconfined criticality scenario of the QPT to the N\'eel phase. A gapless spin-liquid state is also inconsistent with this finding. In accordance with previous results, we observe the first-order QPT at $J_2\approx 0.64$ to another N\'eel phase. Ground state energies we obtain at $0.4\alt J_2 < 0.64$ are in good agreement with results of previous numerical calculations.

The rest of the present paper is organized as follows. We describe our approach in Sec.~\ref{method}. Ground state properties and singlet excitations are discussed in Sec.~\ref{analysis}. Summary and conclusion can be found in Sec.~\ref{conc}. One appendix is added with details of the analysis.

\section{Method and technique}
\label{method}

\begin{figure}
\includegraphics[scale=0.40]{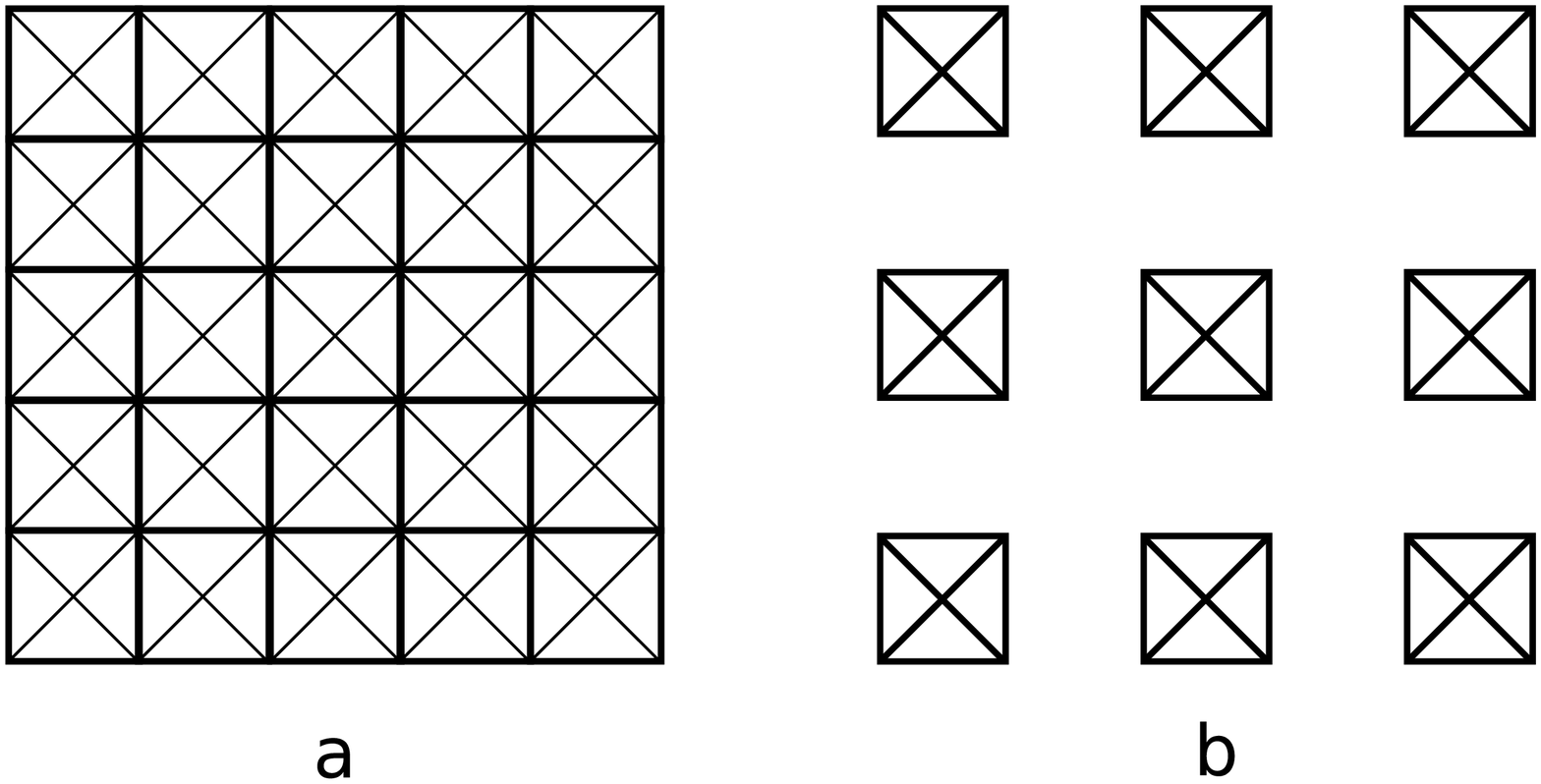}
\caption{
Model \eqref{ham} on simple square lattice (a) and decoupled plaquettes having doubly degenerate singlet ground state (b). Bold and thin lines denote exchange interactions with coupling constants $J_1=1$ and $J_2$, respectively. To come from decoupled plaquettes to model \eqref{ham}, we introduce operator \eqref{v} controlled by parameter $\lambda$ so that $\lambda=1$ and $\lambda=0$ correspond to panels (a) and (b), respectively.
\label{lattice}}
\end{figure}

\begin{figure}
\includegraphics[scale=0.40]{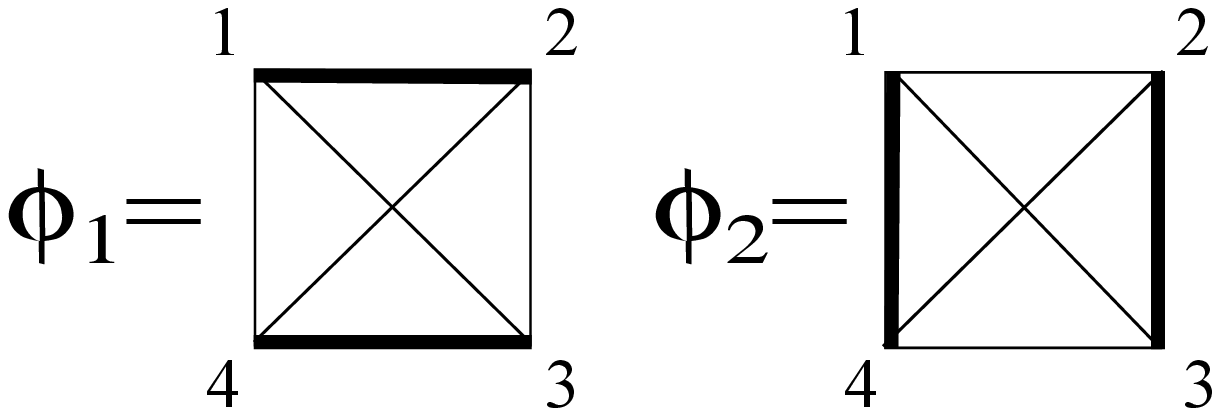}
\caption{ 
Wave functions of an isolated plaquette from which its singlet ground state wave functions \eqref{gswf} are constructed. Bold lines denote singlet states of the corresponding two spins.
\label{bonds}}
\end{figure}

As the method we use is discussed in detail in our previous paper \cite{singlon}  devoted solely to singlet excitations at $J_2=0$, we describe it only briefly here. We perform a sort of plaquette expansion to derive an operator $H$ (an ``effective Hamiltonian'') whose eigenvalues give energies of low-lying singlet levels of model \eqref{ham}. Our starting point is a set of isolated plaquettes in which exchange coupling constants are equal to unity between all four spins (see Fig.~\ref{lattice}(b)). Wave functions of the doubly degenerate singlet ground state of an isolated plaquette
\begin{equation}
\label{gswf}
\Psi^+ = \frac{\phi_1+\phi_2}{\sqrt{3}},
\qquad
\Psi^- = \phi_1-\phi_2
\end{equation}
are constructed as linear combinations of nonorthogonal ones
$
\phi_1=(|\uparrow\rangle_1|\downarrow\rangle_2 
- 
|\downarrow\rangle_1|\uparrow\rangle_2)
(|\uparrow\rangle_3|\downarrow\rangle_4
-
|\downarrow\rangle_3|\uparrow\rangle_4)/2
$
and
$
\phi_2=(|\uparrow\rangle_2|\downarrow\rangle_3 
- 
|\downarrow\rangle_2|\uparrow\rangle_3)
(|\uparrow\rangle_4|\downarrow\rangle_1
-
|\downarrow\rangle_4|\uparrow\rangle_1)/2
$
which are depicted in Fig.~\ref{bonds}. It is convenient to consider $\Psi^+$ and $\Psi^-$ as states of a pseudospin $s=1/2$ corresponding to $s_z=1/2$ and $s_z=-1/2$, respectively.

To come from decoupled plaquettes to model \eqref{ham}, we introduce an operator controlled by a single parameter $\lambda$ (at a given $J_2$) which describes interactions between spins from different plaquettes and which weakens interactions between diagonal spins in each plaquette
\begin{eqnarray}
\label{v}
V &=& 
\lambda 
\left( \sum_{\langle i,j\rangle} 
\left({\bf S}^{(i)}_2{\bf S}^{(j)}_1+{\bf S}^{(i)}_3{\bf S}^{(j)}_4
+
J_2
\left({\bf S}^{(i)}_2{\bf S}^{(j)}_4+{\bf S}^{(i)}_3{\bf S}^{(j)}_1\right)
\right)\right.\nonumber\\
&&{}+
\sum_{\langle i,p\rangle} 
\left({\bf S}^{(i)}_1{\bf S}^{(p)}_4+{\bf S}^{(i)}_2{\bf S}^{(p)}_3
+
J_2
\left({\bf S}^{(i)}_1{\bf S}^{(p)}_3+{\bf S}^{(i)}_2{\bf S}^{(p)}_4\right)
\right)\nonumber\\
&&\left.
{}+(J_2-1)
\sum_i \left({\bf S}^{(i)}_1{\bf S}^{(i)}_3+{\bf S}^{(i)}_2{\bf S}^{(i)}_4\right)
\right),
\end{eqnarray}
where upper and lower indexes of $\bf S$ enumerate plaquettes and the spin number in a plaquette (according to Fig.~\ref{bonds}), respectively, and $\langle i,j\rangle$ and $\langle i,p\rangle$ denote nearest neighbor plaquettes in horizontal and vertical directions, correspondingly. 
Decoupled plaquettes and model \eqref{ham} correspond to $\lambda=0$ and $\lambda=1$, respectively. 

The singlet ground state is $2^N$ times degenerate in the system containing $N$ decoupled plaquettes. This degenerate energy level splits into a singlet band (at $N\to\infty$) upon $\lambda$ increasing because $V$ commutes with operator of the total spin. Our goal is to describe this band in terms of interaction between pseudospins by performing perturbation calculations up to 7-th order in $\lambda$ and extrapolating results to $\lambda=1$ by means of standard methods developed for critical phenomena. We use the standard perturbation theory for systems with degenerate energy levels developed by Bloch and described, e.g., in the textbook \cite{messi}. Consideration of the corresponding wave functions in the spin space is out of the scope of the present paper. We present symmetry arguments in Ref.~\cite{singlon} that it is the low-energy singlet sector of model \eqref{ham} that we consider in this way: no significant overlap is expected of higher-energy singlet bands and the considered one due to the dependence of $V$ on only one control parameter $\lambda$ and different symmetry properties of $\Psi^+$ and $\Psi^-$.

\section{``Effective Hamiltonian'' analysis}
\label{analysis}

The part of $H$ containing terms describing interaction between no more than two pseudospins has the form
\begin{eqnarray}
\label{efh}
H &=& CN + h\sum_i s_i^z
+ \sum_{i,j} 
\left(J^{zz}_{ij}s_i^z s_j^z
+ J^{+z}_{ij}s_i^+ s_j^z
+ J^{-z}_{ij}s_i^- s_j^z
+ J^{z+}_{ij}s_i^z s_j^+
+ J^{z-}_{ij}s_i^z s_j^-
+ J^{++}_{ij}s_i^+ s_j^+
\right.\nonumber\\
&&{}\left.
+ J^{--}_{ij}s_i^- s_j^-
+ J^{-+}_{ij}s_i^- s_j^+
+ J^{+-}_{ij}s_i^+ s_j^-\right),
\end{eqnarray}
where $i$ and $j$ are not restricted to only nearest neighbor plaquettes: terms in the perturbation theory of order higher than three contain two-pseudospin long-range interactions as well as multi-pseudospin interactions. All of them are taken into account in our quantitative consideration below. It should be noted that in accordance with the general property of the perturbation series, \cite{messi} operator $H$ is non-Hermitian (e.g., $J^{++}\ne J^{--}$): non-Hermitian terms appear starting from the third order in $\lambda$ (see Appendix~\ref{app} for series of some coefficients in $H$). However according to a general proof \cite{messi}, its spectrum must be real. Operator $H$ is translationally invariant: it is defined on the square lattice which period twice as large as the period of the original lattice. We calculate corrections to all parameters in $H$ (including the multi-pseudospin and the long-range terms) up to the 7-th order in $\lambda$. To extrapolate the ground-state energy and the spectrum from $\lambda\ll1$ to $\lambda=1$, we use Pad\'e and Pad\'e-Borel resummation techniques. There is no point in applying these techniques individually to each coefficient in $H$ because the number of available terms is small in the series for some of them. Then, we derive below analytical expressions for physical quantities and find series for them up to the 7-th order in $\lambda$ using series for coefficients in $H$.

\subsection{Ground state}

It is seen from Eq.~\eqref{efh} that $H$ describes an anisotropic magnet in effective magnetic field $h$. The magnetic field dominates in $H$ at $\lambda\ll1$ and $J_2\alt0.6$ because the first nonzero term in its series is of the first order in $\lambda$ whereas the first nonzero terms are of the second order in series for coefficients of two-pseudospin interaction (see Appendix~\ref{app} for an example). Then, in the first order in $\lambda$, the system is equivalent to a set of isolated spins in external magnetic field so that $\langle {\bf s}_i\rangle$ is directed along $z$ axis of the pseudospin space. In the second order in $\lambda$, Hermitian two-pseudospin interaction arises in Eq.~\eqref{efh}: $J_{ij}^{++}=J_{ij}^{--}=J_{ij}^{+-}=J_{ij}^{-+}$ and $J_{ij}^{+z}=J_{ij}^{-z}=J_{ij}^{z+}=J_{ij}^{z-}$. As a result $H=\widetilde H$ in the second order, where
\begin{equation}
\label{ht}
	\widetilde H = CN + h\sum_i s_i^z + \sum_{i,j} (J^{zz}_{ij}s_i^z s_j^z + J^{xx}_{ij}s_i^x s_j^x + J^{xz}_{ij} (s_i^x s_j^z + s_i^z s_j^x ) ), 
\end{equation}
$J^{zz}_{ij},J^{xx}_{ij}<0$ and $J^{xz}_{ij}$ have equal moduli and opposite signs on vertical and horizontal bonds (the latter feature originates from different symmetry of $\Psi^+$ and $\Psi^-$: $\Psi^+$ does not change upon plaquette rotation by angle $\pi/2$ whereas $\Psi^-$ changes its sign). Consequently, $\langle {\bf s}_i\rangle$ lie in $xz$ plane of the pseudospin space and, due to inequivalent vertical and horizontal bonds, one has to assume a columnar structure of the ground state shown in Fig.~\ref{gs}(a) and \ref{gs}(b) for $J^{xz}_{ij}>0$ and $J^{xz}_{ij}<0$ on vertical bonds, respectively. These structures are characterized by two angles $\alpha$ and $\beta$ determining directions of $\langle {\bf s}_i\rangle$ with respect to $z$ axis. Notice that one can obtain both $J^{xz}_{ij}>0$ and $J^{xz}_{ij}<0$ on vertical bonds at a given $J_2$ using two bases: wave functions are built in the first basis using $\Psi^+$ and $\Psi^-$ given by Eqs.~\eqref{gswf} whereas $\Psi^+$ and $-\Psi^-$ are used instead in the second basis (i.e., the second basis is obtained from the first one as a result of rotation by angle $\pi/2$). This circumstance leads to an additional ground state degeneracy when $\alpha\ne0$ and/or $\beta\ne0$. For definiteness, we build bare wave functions using Eqs.~\eqref{gswf}. We find as a result of particular calculations that the ground state wave function has the form presented in Fig.~\ref{gs}(a) in this basis when $\alpha\ne0$ and $\beta\ne0$.

At small $\lambda$, when the ``Zeeman term'' dominates in $H$, $\alpha=\beta=0$. However the angles become finite for some $J_2$ at $\lambda\sim1$. We find that no improvement is required of the two-angle ansatz after taking into account other terms (including non-Hermitian ones) in $H$ of higher orders in $\lambda$. In particular, coefficients in series of $J_{ij}^{\pm z}$ and $J_{ij}^{\pm z}$ have equal moduli and opposite signs on vertical and horizontal bonds in all orders in $\lambda$ (see Appendix~\ref{app} for an example).

To obtain a Bose-analog of $H$ and find $\alpha$, $\beta$, and the spectrum, it is natural to use the Holstein-Primakoff transformation. We proceed with the non-Hermitian Bose-analog of $H$ as with a non-Hermitian Bose-analog of a spin Hamiltonian after Dyson-Maleev transformation in magnetically ordered phases, as it was done in Ref.~\cite{singlon}. To obtain the ground state energy $E_{GS}$ at a given $J_2$, we examine terms in $H$ not containing Bose-operators which give a series in $\lambda$ whose coefficients depend on $\alpha$ and $\beta$. Performing the resummation procedure for different values of these angles (spread on a regular grid), we find their values at which the energy is minimal. As it is done in Ref.~\cite{singlon}, we take into account also first $1/s$ corrections to the ground state energy. While their contribution is quite small, they move $E_{GS}$ closer to previous numerical results. 

\begin{figure}
\includegraphics[scale=0.38]{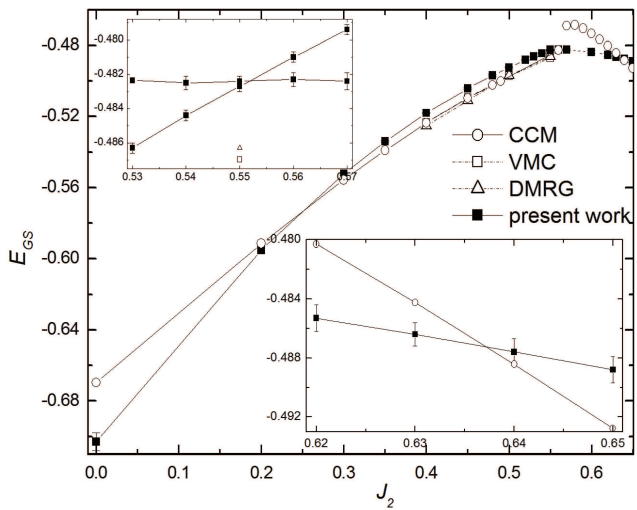}
\caption{ Ground state energy per spin $E_{GS}$ obtained using different approaches: coupled cluster method (CCM), \cite{darradi,richter} variational Monte-Carlo (VMC) calculation with additional Lanczos improvement steps, \cite{vmc} density matrix renormalization group (DMRG) method, \cite{fisher} and the plaquette expansion carried out in the present paper. See also Table~\ref{gsetab} for particular values of $E_{GS}$. Insets illustrate first-order transitions at $J_2\approx0.55$ and $J_2\approx0.64$ as a result of the corresponding level crossing.
\label{gsefig}}
\end{figure}

\begin{figure}
\includegraphics[scale=0.5]{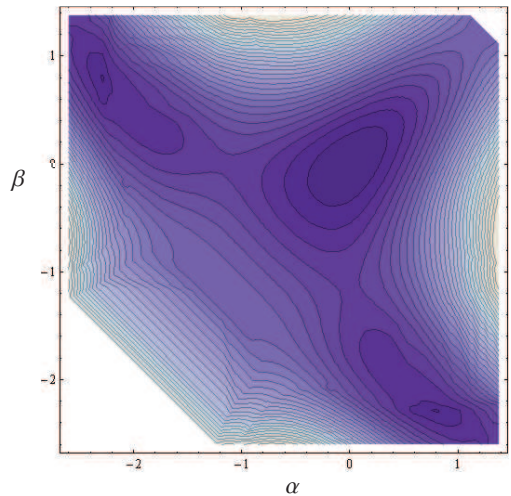}
\caption{(Color online.) Ground state energy per spin as a function of angles $\alpha$ and $\beta$ at $J_2=0.55$ near local minima. Contours connect points with the same energy and points in darker regions have lower energies. The energy difference between points on neighboring contours is $0.0011\cal N$, where $\cal N$ is the number of spins in system \eqref{ham}. The minimum at $\alpha=\beta=0$ and two equivalent minima at $\alpha\approx-2.2$, $\beta\approx0.8$ and $\alpha\approx0.8$, $\beta\approx-2.2$ have approximately the same energy. 
\label{gsland}}
\end{figure}

At $0\le J_2<0.55$, the absolute minimum of the energy is found at $\alpha=\beta=0$ so that the ground state is non-degenerate in the effective model. Obtained values of $E_{GS}$ are presented in Fig.~\ref{gsefig} and in Table~\ref{gsetab}. It is seen that our results are in good agreement with previous numerical findings at $0.2<J_2<0.55$. Quite expectedly, our approach underestimates significantly the ground state energy deep in the N\'eel phase at $J_2<0.2$. 
\footnote{
A second-order phase transition takes place on the way from $\lambda=0$ to $\lambda=1$ to magnetically ordered phase at $J_2\alt0.4$ that leads to a bad convergence of dimer and plaquette expansions at $\lambda\sim1$ (see discussion in Ref.~\cite{singlon}). 
}
Two equivalent local minima (at $\alpha\approx-2.2$, $\beta\approx0.8$ and $\alpha\approx0.8$, $\beta\approx-2.2$) arise at $J_2\approx0.5$. Unlike the minimum at $\alpha=\beta=0$ which energy rises upon $J_2$ increasing, the energy of these local minima is almost independent of $J_2$. Energy of the minimum at $\alpha=\beta=0$ becomes equal at $J_2\approx0.55$ to that of the two equivalent minima signifying a first-order phase transition (see the upper inset in Fig.~\ref{gsefig}). 
\footnote{
In particular, we find at $J_2=0.55$ the following series for the energy at $\alpha=\beta=0$ and $\alpha\approx0.8$, $\beta\approx-2.2$:
$-0.375 - 0.05625 \lambda - 0.0530599 \lambda^2 + 0.00645224 \lambda^3 - 
 0.00679466 \lambda^4 + 0.00411219 \lambda^5 - 0.00415998 \lambda^6 + 0.00778856 \lambda^7$
and
$-0.375 + 0.0304735 \lambda - 0.0952434 \lambda^2 - 0.0203372 \lambda^3 - 
 0.00943256 \lambda^4 - 0.00518608 \lambda^5 - 0.00292147 \lambda^6 - 0.00853038 \lambda^7$, 
respectively. These series give as a result of resummation procedure $-0.4827(3)$ and $-0.4824(3)$, correspondingly. 
}
The energy landscape at $J_2=0.55$ is presented in Fig.~\ref{gsland}. 
\footnote{
Interestingly, one can obtain a picture similar to Fig.~\ref{gsland} in model \eqref{ht}, e.g., at $|J^{xz}|\sim |J^{xx}|\sim |J^{zz}|$ with negative $J^{xx}$ and $J^{zz}$ and an appropriate $h$ value. Our analysis of series for these three coefficients $J$ shows that these relations between them are likely to be valid. Then, the reduced variant of the ``effective Hamiltonian'' \eqref{ht} gives qualitatively correct results.
}
It is seen that all minima are very shallow and barriers between them are low. 
\footnote{
The convergence of series is much better near minima than in the region between them: the estimated error near all minima is $3\cdot10^{-4}\cal N$ ($\cal N$ is the number of spins in the system) while it is an order of magnitude larger between the minima. Then, we can estimate only roughly the barrier height between minima as $0.004(2)\cal N$.
}
It would signify that there are many low-energy states in the system with nearly equal energies at $J_2\sim0.5$.  

As soon as we do not examine ground-state wave functions in the spin space in the present study, it is difficult to say definitely what phases correspond to the energy minima at $J_2\approx0.55$. It seems unlikely, however, that the minimum at $\alpha=\beta=0$ corresponds to the N\'eel phase at $J_2\approx0.55$ because the majority of previous considerations reports the N\'eel phase stability at $J_2<0.4\div0.5$ (the most wide range of $J_2<0.532$ of the N\'eel phase stability is obtained in Ref.~\cite{tensnet}). Bearing in mind that the majority of previous studies report the plaquette VBS phase at $J_2>0.5$, it looks reasonable to suppose that the minimum at $\alpha=\beta=0$ corresponds (at $J_2\approx0.55$) to the plaquette VBS state while a columnar plaquette VBS phase (with an additional twofold degeneracy of the ground state) corresponds to two equivalent minima (see Fig.~\ref{gs}). However further consideration is needed to determine properties of these two phases precisely. Because we do not obtain the singlet gap closure at $J_2<0.64$ (see below), a transition is expected to the N\'eel phase upon the triplet gap closure at $J_2<0.55$ (the N\'eel phase would also correspond to the minimum at $\alpha=\beta=0$).

At $J_2>0.55$, another local minimum arises at $\alpha,\beta\sim4$ which definitely becomes the absolute minimum at $J_2=J_{c2}>0.6$ signifying another first-order QPT. However the convergence of series at that region of the phase space is very bad (presumably because the decoupled plaquettes are too bad starting point for description of the N\'eel phase at $J_2>0.6$). Then, for an accurate determination of $J_{c2}$, we have to compare the energy of the local minimum at $\alpha\approx0.8$, $\beta\approx-2.2$ with the energy of the N\'eel phase found before numerically using another approach (we choose data from Refs.~\cite{darradi,richter} obtained by the coupled cluster method). As it is seen from the lower inset of Fig.~\ref{gsefig}, the first-order transition takes place at $J_2\approx0.64$ in agreement with many previous results (see, e.g., Refs.~\cite{serplaq,tensnet,serdim2,tensnet2,balents,fisher,mf}).

\begin{table}
\caption{
Values of the ground state energy per spin for some $J_2$ which are also shown in Fig.~\ref{gsefig}. Notice that VMC and DMRG data for $J_2=0.4$ and 0.45 are available for finite systems only. The procedure of extrapolation to thermodynamical limit is expected to give slightly higher energies. \cite{fisher,vmc}
\label{gsetab} }
\begin{ruledtabular}
\begin{tabular}{|c|cccc|}
$J_2$ & plaquette expansion & VMC $^a$ & DMRG & CCM $^c$ \\
\hline
0.4 & $-0.5180(3)$ & $-0.52333$ & $-0.5253$ $^b$ & $-0.52354(4)$\\
0.45 & $-0.5042(3)$ & $-0.5094$ & $-0.5110$ $^b$ & $-0.50964(12)$\\
0.5 & $-0.4924(3)$ & $-0.49717$ & $-0.4968$ $^c$ & $-0.4984(2)$\\
0.55 & $-0.4827(3)$ & $-0.48698$ & $-0.4863$ $^c$ & ---
\end{tabular}
$^a$ Ground state energies per spin in systems $L\times L$ with $L=18$.\\
$^b$ Ground state energies per spin in systems with $L=10$.\\
$^c$ Values extrapolated to thermodynamical limit $L\to\infty$.
\end{ruledtabular}
\end{table}

\subsection{Singlet excitations}

\begin{figure}
\includegraphics[scale=0.4]{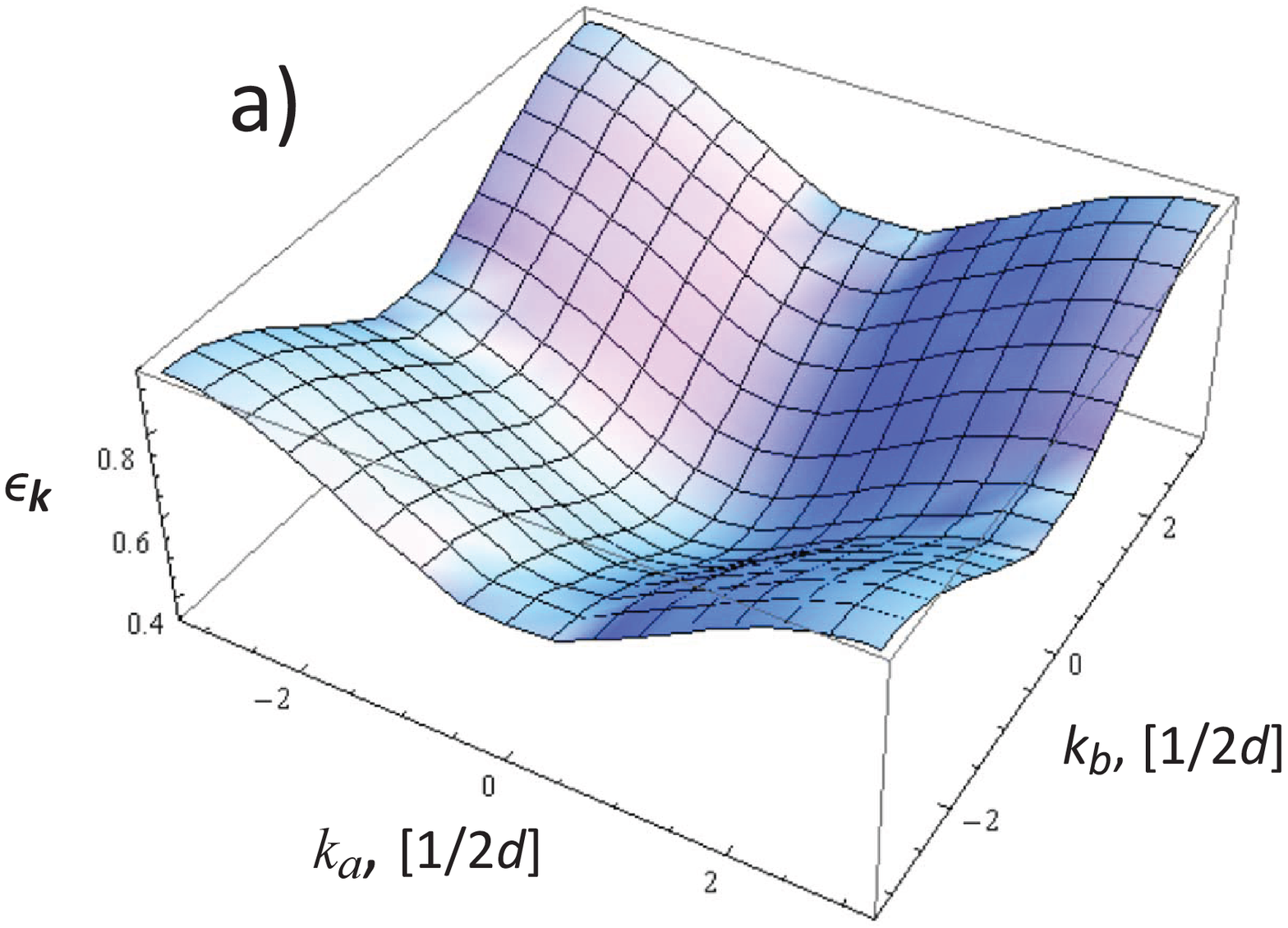}
\includegraphics[scale=0.4]{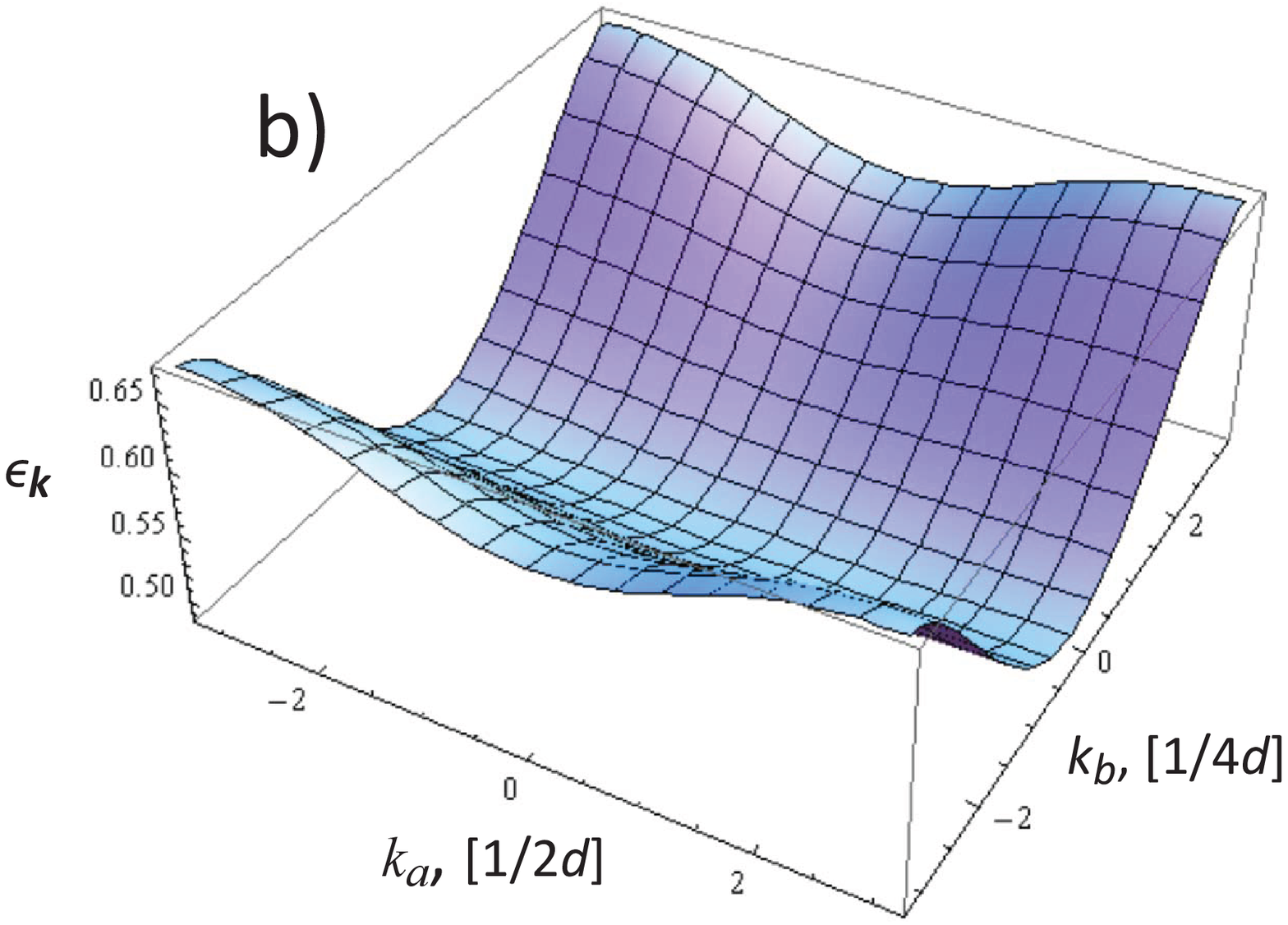}
\caption{(Color online.) Spectra of singlet excitations $\epsilon_{\bf k}$ at $J_2=0.55$ corresponding to (a) ground state energy minimum at $\alpha=\beta=0$ and (b) that at $\alpha\approx0.8$, $\beta\approx-2.2$ (see Fig.~\ref{gsland}). Here $d$ is the distance between nearest-neighbor spins in model \eqref{ham} and $k_{a,b}$ are components of $\bf k$ in the coordinate system shown in Fig.~\ref{gs}(a).
\label{specfig}}
\end{figure}

\begin{figure}
\includegraphics[scale=0.35]{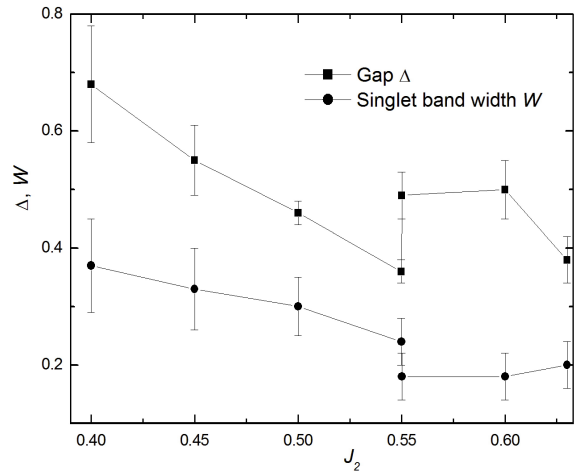}
\caption{ Gap $\Delta$ in the singlet spectrum and the low-energy singlet band width $W$. 
\label{gapfig}}
\end{figure}

Spectrum of singlet excitations is found by analysis of terms in $H$ bilinear in Bose-operators, as it is done in Ref.~\cite{singlon}. Singlet spectra at $J_2=0.55$ are shown in Fig.~\ref{specfig} corresponding to minima at $\alpha=\beta=0$ and at $\alpha\approx0.8$, $\beta\approx-2.2$. Remarkably, singlet spectra have quite large gaps at $J_2<0.64$ which values are presented in Fig.~\ref{gapfig}. At $J_2<0.55$, the gap is located at $\bf k=0$. Small dispersion of the spectrum along $a$-axis (see Fig.~\ref{gs}(a)) at $J_2>0.55$ does not allow to determine the gap location on the line $(-\pi,0)$--$(\pi,0)$. Estimation of the first $1/s$ corrections to the spectrum demonstrates a very small renormalization by quantum fluctuations of pseudospins.

\section{Conclusion}
\label{conc}

To conclude, we construct operator $H$ whose eigenvalues coincide with energies of low-energy singlet states of spin-$\frac12$ $J_1$--$J_2$ Heisenberg model \eqref{ham} on square lattice. Parameters of $H$ are found in the first seven orders of the perturbation theory. Using series for coefficients, we obtain series for the ground state energy and the singlet spectrum. It is expected that our approach is particularly suitable for discussion of the plaquette phase observed in many previous papers at $0.4\alt J_2\alt0.6$. After standard resummation procedure, we obtain the ground state energies presented in Fig.~\ref{gsefig} and in Table~\ref{gsetab}. In accordance with results of previous considerations, we find first-order transition from the magnetically disordered phase to the N\'eel one at $J_2\approx0.64$. We observe also a first-order QPT at $J_2\approx0.55$. Bearing in mind results of previous considerations, it seems unlikely that this QPT separates another N\'eel phase and the non-magnetic one (because we analyze only singlet spectrum in the present study, we cannot make a definite conclusion about properties of phases). We suppose that this is a QPT inside the non-magnetic region between the plaquette phase (stable at $J_2<0.55$) and the phase arisen at $0.55<J_2<0.64$ which has presumably the columnar plaquette structure with additional twofold ground state degeneracy related with the rotational symmetry breaking (see Fig.~\ref{gs}). However, further consideration is needed to determine properties of these two phases precisely. 

The singlet spectrum is found to be gapped at $J_2<0.64$ (see Fig.~\ref{gapfig}) that makes impossible neither a spin-liquid state in the non-magnetic region no the deconfined quantum criticality scenario for the transition to the N\'eel phase at $J_2<0.55$. The latter transition would be characterized by a closure of only the triplet gap.

\begin{acknowledgments}

We thank J.\ Richter for exchange of data. This work is supported by Russian Science Foundation (grant No.\ 14-22-00281).

\end{acknowledgments}

\appendix

\section{Coefficients of the ``effective Hamiltonian'' at $J_2=0.55$}
\label{app}

Coefficients of the ``effective Hamiltonian'' \eqref{efh} are shown in Table~\ref{coef} in the first seven orders in $\lambda$ describing interaction between nearest and next-nearest neighbor pseudospins at $J_2=0.55$. Numerous coefficients for long-range and multi-pseudospin interactions have been also calculated. They are also taken into account in the quantitative analysis carried out in the present paper and they can be provided upon request.

\begin{table}
\caption{
Nonzero coefficients of the ``effective Hamiltonian'' in the first seven orders in $\lambda$ describing interaction between nearest and next-nearest neighbor pseudospins at $J_2=0.55$. Subscripts $h$, $v$, and $d$ denote the shortest horizontal, vertical, and diagonal bonds between pseudospins, respectively. Constant $C$ has also the zero-order correction equal to $-\lambda^0 3/2$ which is given by the ground state energy of an isolated plaquette.
\label{coef}
}
\begin{ruledtabular}
\begin{tabular}{cccccccc}
& 1&2&3&4&5&6&7\\
\hline
$C$ &  0.225 & $-$0.324427 & $-$0.061202 & $-$0.050632 & $-$0.010787 & $-$0.028230 & $-$0.009198\\
$h$ & 0.9 & $-$0.332917 & $-$0.277432 & $-$0.105482 & $-$0.102216 & $-$0.037882 & $-$0.110016\\
\hline
$J_h^{zz}$ &0 & $-$0.108542 & $-$0.085331 & $-$0.035047 & $-$0.033701 & $-$0.008582 & $-$0.028041\\
$J_h^{--}$ &0 & $-$0.081406 & 0.044564 & 0.042000 & 0.097999 & 0.034800 & 0.146126\\
$J_h^{++}$ &0 & $-$0.081406 & $-$0.028701 & $-$0.073196 & $-$0.073308 & $-$0.090835 & $-$0.102988\\
$J_h^{+-}=J_h^{-+}$ &0 & $-$0.081406 & 0.007932 & 0.000887 & 0.011299 & 0.005432 & 0.017833\\
$J_h^{-z}=J_h^{z-}$ &0 & $-$0.094000 & $-$0.011220 & 0.051871 & 0.053753 & 0.014496 & $-$0.089420\\
$J_h^{+z}=J_h^{z+}$ &0 & $-$0.094000 & $-$0.053520 & $-$0.041564 & $-$0.063113 & 0.013452 & 0.076528\\
\hline
$J_d^{zz}$ &0 & 0 & $-$0.013664 & $-$0.017605 & 0.004330 & $-$0.001672 & 0.020554\\
$J_d^{--}$ &0 & 0 & 0.010248 & 0.009056 & 0.044433 & 0.027467 & 0.041915\\
$J_d^{++}$ &0 & 0 & 0.010248 & $-$0.009390 & $-$0.039010 & $-$0.042935 & $-$0.042100\\
$J_d^{+-}=J_d^{-+}$ &0 & 0 & 0.010248 & $-$0.000167 & $-$0.004437 & $-$0.003411 & $-$0.006107\\
\hline
$J_v^{zz}$ &0 & $-$0.108542 & $-$0.085331 & $-$0.035047 & $-$0.033701 & $-$0.008582 & $-$0.028041\\
$J_v^{--}$ &0 & $-$0.081406 & 0.044564 & 0.042000 & 0.097999 & 0.034800 & 0.146126\\
$J_v^{++}$ &0 & $-$0.081406 & $-$0.028701 & $-$0.073196 & $-$0.073308 & $-$0.090835 & $-$0.102988\\
$J_v^{+-}=J_v^{-+}$ &0 & $-$0.081406 & 0.007932 & 0.000887 & 0.011299 & 0.005432 & 0.017833\\
$J_v^{-z}=J_v^{z-}$ &0 & 0.094000 & 0.011220 & $-$0.051871 & $-$0.053753 & $-$0.014496 & 0.089420\\
$J_v^{+z}=J_v^{z+}$ &0 & 0.094000 & 0.053520 & 0.041564 & 0.063113 & $-$0.013452 & $-$0.076528
\end{tabular}
\end{ruledtabular}
\end{table}

\bibliography{j1j2bib}

\end{document}